\documentclass{aa}
\usepackage[utf8]{inputenc}
\usepackage{graphicx}
\usepackage{txfonts}
\usepackage{mathtools}
\usepackage{nameref}
\usepackage{wasysym}
\usepackage{makecell}
\usepackage[caption=false]{subfig}
\usepackage{hyperref}
\usepackage{xcolor}
\usepackage{float}
\usepackage{booktabs}
\usepackage[capitalise]{cleveref}
\usepackage{natbib}
\usepackage{soul}

\newcommand{\addone}[1]{#1}
\newcommand{\editone}[2]{#2}

\graphicspath{{img/}}

\hypersetup{
    colorlinks=true,
    linkcolor=blue,
    citecolor=blue,
    filecolor=magenta,      
    urlcolor=blue,
}
\bibpunct{(}{)}{;}{a}{}{,} % to follow the A&A style
\begin{document}

\author{Simon Müller\inst{1}
        \and
        Ravit Helled\inst{1}
       }
\authorrunning{Müller \& Helled}

\title{Towards a new era in giant exoplanet characterisation}

\institute{Center for Theoretical Astrophysics and Cosmology \\
           Institute for Computational Science, University of Zürich \\
           Winterthurerstrasse 190, 8057 Zürich, Switzerland \\
          \email{simonandres.mueller@uzh.ch}
          }

\date{Received 28 September 2022 / Accepted 2 November 2022}

\abstract{
Determining the composition of giant exoplanets is crucial for understanding their origin and evolution. However, planetary bulk composition is not measured directly but must be deduced from a combination of mass-radius measurements, knowledge of the planetary age and evolution simulations. Accurate determinations of stellar ages, mass-radius measurements, and atmospheric compositions from upcoming missions can significantly improve the determination of the heavy-element mass in giant planets.
In this paper we first demonstrate the importance of an accurate age measurement, as expected from Plato, in constraining planetary properties. Well-determined stellar ages can reduce the bulk-metallicity uncertainty by up to a factor of two.
We next infer the bulk metallicity of warm giants from the Ariel mission reference sample and identify the Ariel high-priority targets for which a measured atmospheric metallicity can clearly break the degeneracy in the inferred composition. We show that knowledge of the atmospheric metallicity can broadly reduce the bulk-metallicity uncertainty by a factor of four to eight.
We conclude that accurate age determinations from Plato and atmospheric measurements by Ariel and the \textit{James Webb} Space Telescope will play a key role in revealing the composition of giant exoplanets. 
}

\maketitle

\section{Introduction}\label{sec:introduction}

Understanding the origin of giant planets and their characterisation are two key objectives of exoplanetary science. Determining their atmospheric and bulk compositions is particularly important: they can be compared with theoretical predictions and be used to constrain formation and evolution pathways \citep[e.g.][]{Guillot2006,Mousis2009,Johansen2017,Hasegawa2018,Ginzburg2020,Shibata2020,Helled2022}. However, the planetary bulk composition cannot be directly measured and has to be inferred with theoretical models from mass-radius measurements and stellar ages \citep[e.g.][]{Fortney2007,Miller2011,Thorngren2016,Teske2019}. 
The inferred composition depends on model assumptions such as the equations of state \citep{Baraffe2008,Vazan2013}, the atmospheric model \citep{Poser2019} and the assumed atmospheric metallicity \citep{Burrows2007,2020ApJ...903..147M}. This leads to an intrinsic uncertainty in the predicted bulk metallicity since there are several possible solutions for a given set of measured planetary parameters.
Part of this uncertainty can be reduced with more accurate mass and radius measurements expected from ground-based facilities such as High Accuracy Radial velocity Planet Searcher (HARPS; \citet{2003Msngr.114...20M}), Near Infra-Red Planet Searcher (NIRPS; \citet{2017Msngr.169...21B}) and Echelle SPectrograph for Rocky Exoplanets and Stable Spectroscopic Observations (ESPRESSO; \citet{2021A&A...645A..96P}), as well as the upcoming PLAnetary Transits and Oscillations of stars mission (Plato; \citet{Rauer2014}) which will also provide an accurate determination of the stellar age.

Unlike the bulk composition, the atmospheric composition of giant exoplanets is measurable. In recent years, there have been detections of several chemical species in hot Jupiter atmospheres \citep[e.g.][]{2014ApJ...791L...9M,2017ApJ...834...50B,2017Natur.549..238S,2018Natur.557..526N,2019MNRAS.482.1485P,2019ApJ...887L..20W,2020ApJ...897L...5B,2021Natur.598..580L}. However, the atmospheric characterisation of warm Jupiters has so far been limited. The \textit{James Webb} Space Telescope (JWST; \citet{Gardner2006,2015MNRAS.448.2546B}) and in the near future the Ariel mission \citep{Tinetti2018} will change that with high-precision measurements of exoplanet atmospheres.

One reason atmospheric measurements are vital for the characterisation of giant exoplanets is that their inferred bulk compositions are degenerate with respect to the assumed atmospheric metallicity \citep{2020ApJ...903..147M,2021MNRAS.507.2094M}. Therefore, these measurements promise to further constrain the bulk composition and therefore the planetary interiors, and provide crucial information for linking the interiors and atmospheres of giant planets \citep{2019ApJ...874L..31T}.

Clearly, the combination of accurate mass, radius, age and atmospheric composition measurements will launch a new era in exoplanet characterisation. In order to take full advantage of these exciting future data, detailed theoretical investigations are required. Such models can identify conditions and key targets for which the new measurements will yield the greatest gain in knowledge about planetary interiors.

In this work we investigate how the upcoming Plato and Ariel missions can improve our knowledge of the composition of giant exoplanets and suggest high-value targets for observations. The paper is structured as follows: In Sect. \ref{sec:methods} we describe the thermal evolution models and how they are used to estimate the bulk metallicity of giant exoplanets. We then apply this framework to investigate under which conditions an accurate stellar age measurement can help constrain the heavy-element mass (Sect. \ref{sec:connection_to_plato}). In Sect. \ref{sec:connection_to_ariel} we investigate how the determination of the atmospheric composition with the Ariel mission can also independently be used as a further constraint. We determine the planets from the Ariel mission reference sample for which this is the case, and suggest several key targets for further investigation. We calculate the mass-metallicity relation for the planets in the Ariel sample and compare it to values from previous studies in Sect. \ref{sec:mass_metallicity}. 
\addone{In Sect. \ref{sec:collisions} we investigate how collisions could contribute towards inflating giant planets, which could be important for future observations.}
We discuss our findings and the limitations of our models in Sect. \ref{sec:discussion}. Finally, our conclusions are summarised in Sect. \ref{sec:conclusions}.

\section{Methods}\label{sec:methods}

We considered a planet with a measured mass ($M_{obs}$), radius ($R_{obs}$) and host-star age ($\tau_{obs}$). The age of the planet was assumed to be the same as the star. Since the protoplanetary disk dissipates within about a million years \citep{2009AIPC.1158....3M} and the giant planet has to be formed before that, this is generally a valid simplification. The planet's composition is unknown, but we can calculate evolution models for the measured mass with different bulk metallicities ($Z$), until we find the composition for which the prediction from the model matches the observed radius ($R_{model} = R_{obs}$), at time $\tau_{obs}$. This would, in principle, yield a unique bulk metallicity for an observed exoplanet. 

However, there are two complications: First, there are measurement uncertainties. Second, the theoretical models have to make a range of assumptions, for example, the equations of state for the different chemical elements, their distribution in the planetary interior, the atmospheric model and the treatment of incident stellar irradiation. These assumptions lead to additional uncertainties in the predictions \citep{Burrows2007,Baraffe2008,Thorngren2016, 2020ApJ...903..147M}. Therefore, the evolution models predict model-dependent probability distributions of the bulk metallicity $p(Z \, | \, M_{obs}, R_{obs}, \tau_{obs})$ rather than a single value. These distributions are usually well approximated by a Gaussian $p(Z \, | \, M_{obs}, R_{obs}, \tau_{obs}) \sim \mathcal{N}(\mu_Z, \sigma_Z$) with mean value $\mu_Z$ and standard deviation $\sigma_Z$.

In practice, $\mu_Z$ and $\sigma_Z$ are inferred with a Monte Carlo approach by repeatedly drawing a mass ($M$), radius ($R$) and age ( $\tau$) ,from prior distributions. For the mass and radius priors, we assume normal distributions $\mathcal{N}(\mu_{M, R}, \sigma_{M, R})$ with mean values $\mu_{M, R}$ and standard deviations $\sigma_{M, R}$. The age prior is assumed to follow a uniform distribution $\mathcal{U}(\tau_{min}, \tau_{max})$ with lower and upper limits ($\tau_{min}$ and $\tau_{max}$) given by observations of the host stars. Each draw of a set of parameters $(M, R, \tau)$ is used to estimate the bulk metallicity as described above. If repeated sufficiently many times, the posterior distribution $p(Z \, | \, M_{obs}, R_{obs}, \tau_{obs})$ can be estimated.

For an easier comparison between different cases, we often use the relative uncertainty $r_x \equiv \mu_x / x$ instead of the uncertainty $\sigma_x$. The symbols we used throughout this work are listed and described in \cref{tab:symbols}.

\begin{table}[ht]
    \centering
    \setlength{\tabcolsep}{6pt}
    \begin{tabular}{c|c}
        \toprule
        Symbol & Description \\ 
        \midrule
        $\mu_x$     & \makecell{mean value of the \\ measured or inferred parameter} \\
        $\sigma_x$  & \makecell{standard deviation of the \\ measured or inferred parameter} \\
        $r_x$  & \makecell{relative uncertainty of the \\ measured or parameter ($\mu_x / x$)} \\ 
        \bottomrule
    \end{tabular}
    \caption{Definitions used in this work. $x$ is a place-holder variable for the radius ($R$), mass ($M$) stellar age ($\tau$), or the inferred bulk metallicity ($Z$).}
    \label{tab:symbols}
    \setlength{\tabcolsep}{6pt}
\end{table}

To calculate the planetary evolution we used the python package \textit{planetsynth} \citep{2021MNRAS.507.2094M}. Instead of solving the planetary structure and evolution equations \citep[e.g.][]{Kippenhahn2012} directly, \textit{planetsynth} generates synthetic cooling tracks by interpolating on a large grid of pre-calculated thermal evolution models. For our purposes, \textit{planetsynth} calculated the planetary radius $R(t \, | \, M, Z, Z_{atm}, F_*)$ as a function of time $t$ and the planetary parameters mass $M$, bulk metallicity $Z$, atmospheric metallicity $Z_{atm}$ and incident stellar irradiation $F_*$.

Underlying the synthetic cooling tracks are thermal evolution models calculated with a modified version \citep{Mueller2020} of the stellar evolution code Modules for Experiments in Stellar Astrophysics (MESA; \citet{Paxton2011,Paxton2013,Paxton2015,Paxton2018,Paxton2019,2022arXiv220803651J}). As stated above, the output of these simulations depends (among other parameters) on the equation of state and the atmospheric model. \citet{2021MNRAS.507.2094M} used the \citet{Chabrier2019} hydrogen-helium equation of state together with a 50-50 rock-water mixture equation of state for the heavy elements based on the Quotidian Equation of State (QEoS; \citet{More1988,Vazan2013}). The atmospheric model includes the effect of the heavy elements on the radiative opacity \citep{Freedman2014} as well as the incident stellar irradiation. \addone{A limitation of the \citet{Freedman2014} opacity is that the heavy elements are assumed to be present in solar ratios, which is unlikely to be the case for giant planets in general. However, radiative opacities that consider different elemental ratios and can be used in thermal evolution models are currently unavailable. While our fiducial models do not include the effect of grains, we briefly investigate their effect in Appendix \ref{sec:appendix_grain_opacity}.} More details on the evolution models and their equation of state, atmospheric model, and initial conditions can be found in \citet{2021MNRAS.507.2094M}.

\section{The importance of measuring the planetary age and the connection to the Plato mission}\label{sec:connection_to_plato}

In this section we explore how the accurate determination of stellar ages can further constrain the composition of giant exoplanets. In order to guide the target selection for the Plato mission, we also study how this depends on the stellar age and the irradiation that a planet receives. 

We considered a synthetic planet with $R = 1 R_J$, $M = 1 M_J$ and solar atmospheric metallicity that is being irradiated by two different stellar fluxes of either $F_* = 10^5$ (low irradiation) or $F_* = 10^8$ erg cm$^{-2}$ s$^{-1}$ (high irradiation), and three stellar ages of 0.1, 3 and 8 Gyr. We then considered three different relative stellar age uncertainties of $r_\tau = 50, 25$ and $10$\%. For the mass and radius priors, we used a base-line standard deviation corresponding to an observational uncertainty of 1\%. Overall, these combinations of parameters resulted in 18 synthetic planets for which the metallicity was estimated with 1,000,000 samples (for each planet) as described in Sect. \ref{sec:methods}.

\begin{figure}
    \centering
    \includegraphics[width=\columnwidth]{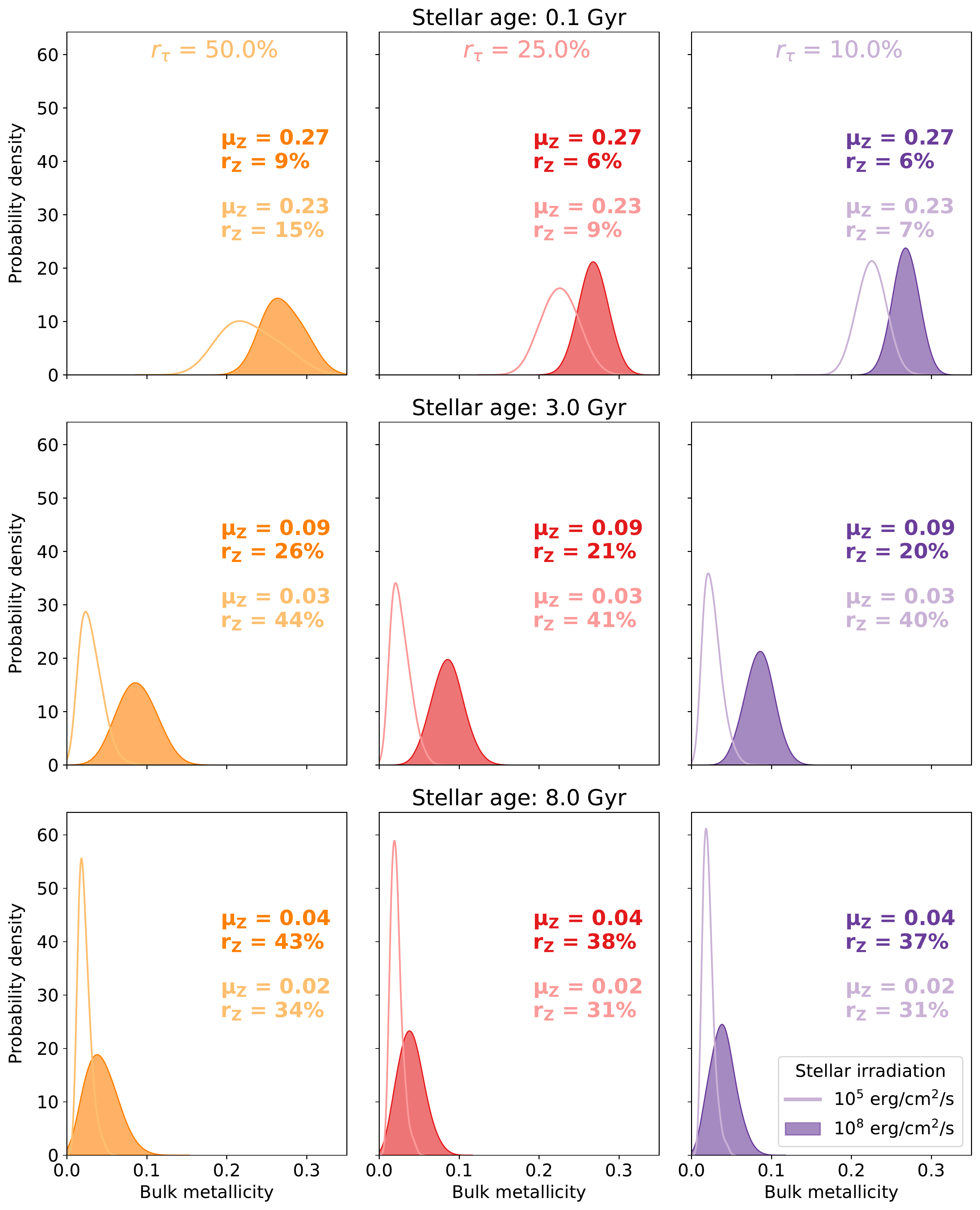}
    \caption{Distributions of the inferred metallicities for a synthetic planet ($R = 1 R_J$ and $M = 1 M_J$) at an age of 0.1 (top row), 3 (middle row), and 8 Gyr (bottom row). The panels in each row show the probability densities for decreasing stellar age uncertainties: $r_\tau = 50 \%$ (left column), $25 \%$ (middle column), and $10 \%$ (right column). The lines show the probability densities for low stellar irradiation, and the filled curves for high irradiation ($F_* = 10^5$ and $10^8$ erg cm$^{-2}$ s$^{-1}$, respectively) Normal and bold font weights correspond to the low and high irradiation models.}
    \label{fig:metallicity_dists_synthetic_R1.0M1.0_grid}
\end{figure}

The resulting probability densities for the bulk metallicities are shown in \cref{fig:metallicity_dists_synthetic_R1.0M1.0_grid}. For very young planets (0.1 Gyr), reducing $r_\tau = 50\%$ to $r_\tau = 10\%$ reduces the relative uncertainty in the metallicity significantly. This depends on whether the irradiation is weak or strong, with the weakly irradiated planets benefiting more from the lower age uncertainty. For intermediate-age and old planets (3 and 8 Gyr) the opposite is true: an accurate stellar age is more important for highly irradiated planets. 

Independent of the irradiation is the fact that older planets have (in absolute terms) more accurately determined compositions compared to younger ones. Also, the age uncertainty has less influence on $r_Z$ for older planets, with $r_\tau = 10\%$ causing $\sigma_Z$ to be smaller by $\sim$10-30\% compared to the $r_\tau = 50\%$ case. This is because the intrinsic luminosity decreases as planets age and therefore their radii only change slowly with time. At early times, the difference in radius for two different times can be much more dramatic.

\begin{figure}
    \centering
    \includegraphics[width=\columnwidth]{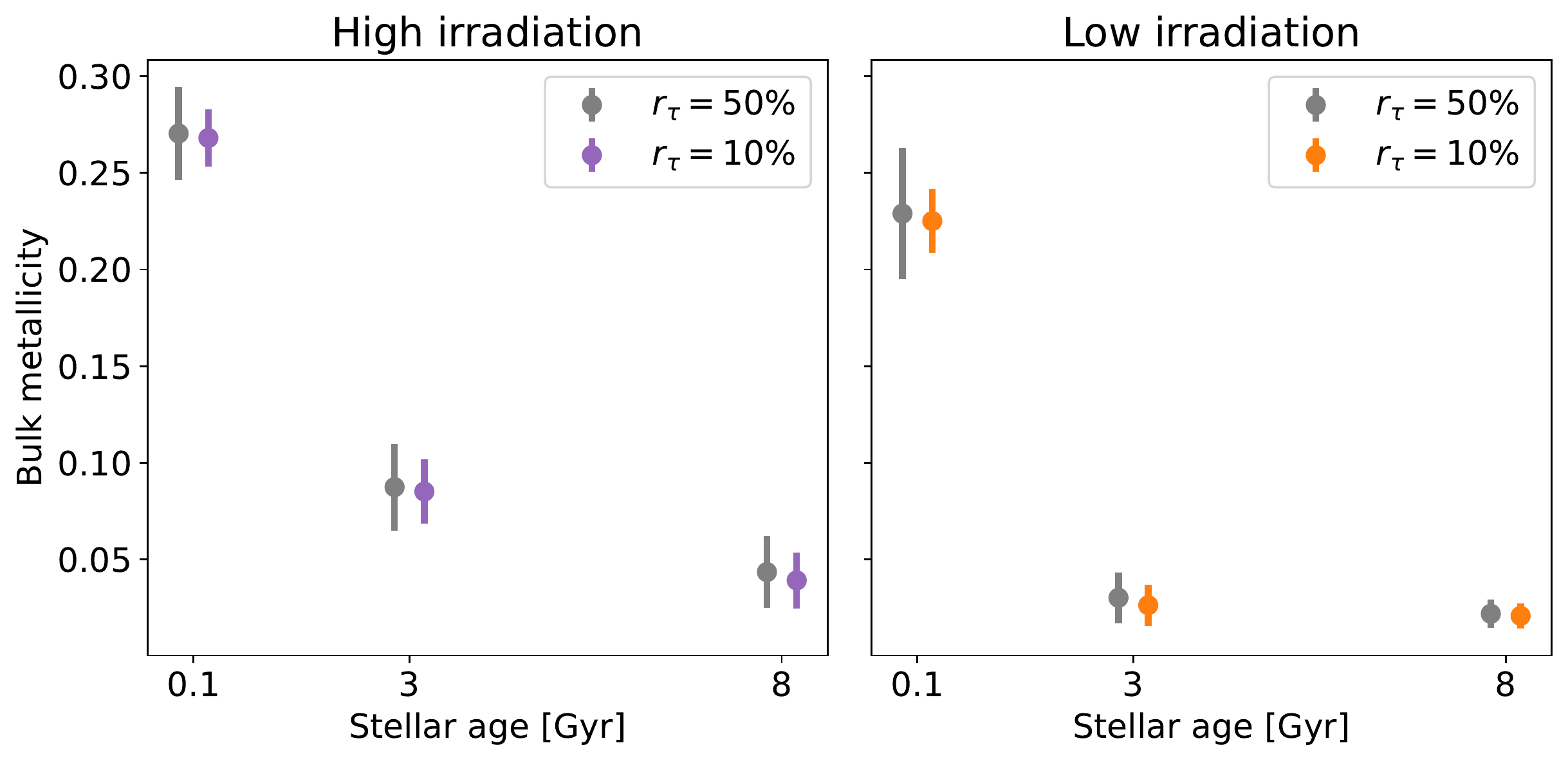}
    \caption{Bulk metallicities and their 1$\sigma$ uncertainties for the synthetic planet with high irradiation (left panel) and low irradiation (right panel). The grey circles and coloured hexagons are for relative age uncertainties of $r_\tau = 50\%$ and $r_\tau = 10\%$, respectively.}
    \label{fig:uncertainty_comparison_synthetic}
\end{figure}

\cref{fig:uncertainty_comparison_synthetic} shows the inferred bulk metallicities and their uncertainties from the models with 50\% and 10\% stellar age uncertainties. This is depicted as a function of age for high (left panel) and low stellar irradiation (right panel). In general, the bulk metallicity for old, weakly irradiated giant planets is better constrained. As discussed above, this only modestly depends on the stellar age uncertainty. However, in particular young and intermediate-aged planets benefit greatly from a low $r_\tau$. Compared to a 50\% uncertainty in stellar age, a 10\% uncertainty decreases the uncertainty in the bulk-metallicity by up to a factor of two. It is therefore clear that accurate stellar age measurements are highly valuable for the characterisation of giant exoplanets.

\section{The importance of atmospheric measurements and the connection to the Ariel mission}\label{sec:connection_to_ariel}

Atmospheric measurements of giant planets can significantly improve our understanding of the planetary bulk composition. First, the atmospheric metallicity can be assumed to be a lower-bound for the bulk ($Z_{atm} \leq Z$). For a homogeneously mixed planet, the atmospheric composition would be the same as the bulk ($Z_{atm} = Z$). However, from interior models of Jupiter and Saturn that match observational constraints we expect deep interiors with a higher metallicity than the envelope \citep[e.g.][]{Wahl2017,Debras2019,2022A&A...662A..18M}. Therefore, the lower-bound assumption is valid as long as the location where the heavy elements are measured is not enriched compared to the rest of the outer envelope. Second, the inferred metallicity from mass-radius measurements is degenerate with respect to the assumed atmospheric composition \citep{2020ApJ...903..147M}. At a given age, the theoretically predicted radius can be significantly different depending on the atmospheric metallicity, because giant planets with enriched atmospheres cool more slowly \citep{Burrows2007}. An atmospheric measurement is therefore an additional and important constraint for the evolution modelling of giant planets.

In this section we demonstrate the importance of atmospheric metallicity measurements and suggest key targets for future observations, focusing on the planets in the Ariel mission reference sample \citep{2022arXiv220505073E}. The sample was reduced to include cool-to-warm giant planets with measured masses, radii and stellar ages. We define cool-to-warm giant planets to have a mass between 0.1 - 10 $M_J$ with an incident stellar irradiation smaller than $F_* \leq 2 \times 10^8$ erg cm$^{-2}$ s$^{-1}$, which is the conventional irradiation flux cutoff for inflated hot Jupiters \citep{Miller2011}. They were excluded since their inflation mechanism is still not well understood \citep{Fortney2010,Weiss2013,Baraffe2014}, and therefore their bulk metallicities cannot be estimated properly.

We identify 27 planets in the Ariel mission reference sample that match these criteria. There are no cool giants (with effective temperatures similar to Jupiter) among these planets, and their semi-major axes are all below 0.3 au. The lowest irradiation flux is received by TOI-1899b with a value of $F_* \simeq 4 \times 10^6$ erg cm$^{-2}$ s$^{-1}$. For comparison, this is still about two \editone{order of magnitudes}{orders of magnitude} higher than the flux that Jupiter receives ($F_* \simeq 5 \times 10^4$ erg cm$^{-2}$ s$^{-1}$). The planets are diverse in terms of their masses, and cover the entire parameter space. There are a few planets with similar masses but rather different radii, suggesting a variety of ages or compositions. Further details on the 27 planets are given in Appendix \ref{sec:potential_ariel_targets}. There, the planets are listed in \cref{fig:planets_sample_listed}, along with their semi-major axis and their incident stellar irradiation. We also show the measured masses and radii of these 27 planets in \cref{fig:planets_sample}.

We inferred the composition of each planet for four different assumed atmospheric metallicities: $Z_{atm} = 0.5, 1, 4, 8 Z_\odot$, where $Z_\odot$ is the solar metallicity. For each $Z_{atm}$ we inferred the metallicity from 1,000,000 samples (see Sect. \ref{sec:methods} for details) constructed from the observed masses, radii, stellar ages and their associated uncertainties\footnote{If a planet had asymmetric lower and upper errors for the mass or radius, we used their arithmetic mean as the standard deviation for Gaussian sampling purposes.}. This resulted in estimates for the bulk-metallicity probability distributions for each planet, allowing us to investigate whether they are overlapping or significantly different. In the latter case, a measurement of $Z_{atm}$ would significantly constrain the bulk composition.

\begin{figure}[ht]
    \centering
    \includegraphics[width=\columnwidth]{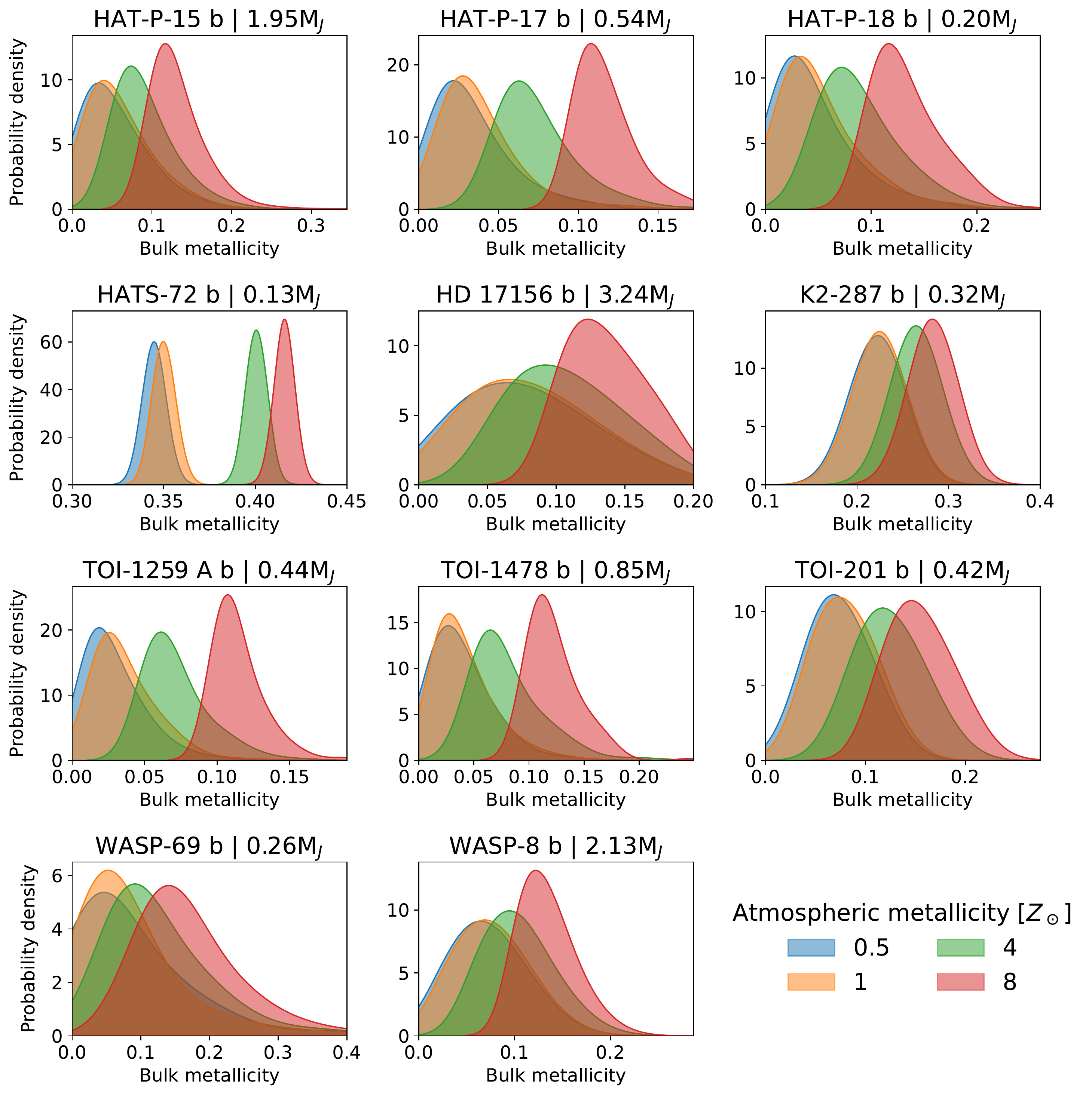}
    \caption{Probability densities of the inferred metallicities for 11 planets from our sample. The colours are the distributions for different atmospheric metallicities (see the legend). The depicted planets show at least moderate separation between the peaks in the different Gaussian distributions (see text for details).}
    \label{fig:metallicity_dists_ariel}
\end{figure}

\cref{fig:metallicity_dists_ariel} shows the probability densities for the inferred bulk metallicities of 11 planets in our sample for each atmospheric metallicity. These planets and their parameters are listed in \cref{tab:key_ariel_targets}. They show a moderate to clear separation of the Gaussian peaks, and are therefore promising targets for atmospheric characterisation by Ariel. For the other 16 planets from the sample, the probability densities for different atmospheric metallicities were moderately to strongly overlapping. As a result, given the current observational uncertainties, an atmospheric measurement would not help constrain their bulk compositions. However, we stress that this may change in the future as planetary parameters are measured more precisely.

There are two planets whose metallicities could not be inferred. First, HAT-P 20b with $R = 0.87 \pm 0.03 R_J$, $M = 7.26 \pm 0.18 M_J$, $2.9 \leq \tau \, \textrm{[Gyr]} \leq 12.4$ and $F_* = 1.99 \times 10^8$ erg cm$^{-2}$ s$^{-1}$ \citep{2011ApJ...742..116B}. This planet is known to be extremely dense and metal-rich, and could not be simulated with our models due to equation of state limitations. Previous estimates suggest a very high total heavy-element mass of $M_Z \simeq 660 \, M_\oplus$ \citep{Thorngren2016}. We note that a more recent measurement \citep{2017A&A...601A..53E} inferred $R = 1.03 \pm 0.05 R_J$ at a similar mass, in disagreement with the radius given by \citet{2011ApJ...742..116B}. This alternative is consistent with a solar or few times solar \editone{composition}{metallicity}. 

Second, TOI-1899 b has $R = 1.15 \pm 0.05 R_J$, $M = 0.66 \pm 0.07 M_J$, $2.7 \leq \tau \, \textrm{[Gyr]} \leq 11.8,$ and $F_* = 3.95 \times 10^6$ erg cm$^{-2}$ s$^{-1}$ \citep{2020AJ....160..147C}. For its age, the measured radius is inflated by $\sim$15\% compared to predictions from the evolution model, despite the irradiation flux being two orders of magnitude below the hot Jupiter cutoff. Giant exoplanets with very high or very low densities such as HAT-P 20b and TOI-1899 b demonstrate the need for more detailed evolution simulations, as well as additional observations to confirm and further constrain their masses and radii. 

There are a few general trends to be noted. First, a higher atmospheric metallicity increases the inferred bulk metallicity. This is because a metal-enriched atmosphere is more opaque, which delays the cooling and leads to a larger radius at a given age \citep{Burrows2007,Mueller2020}. Second, for some planets only the low $Z_{atm} = 0.5, 1 Z_\odot$ and high $Z_{atm} = 8 Z_\odot$ cases were clearly separated, with the intermediate $Z_{atm} = 4 Z_\odot$ overlapping the two distributions by various degrees.

There are, however, planets for which the peaks of the distributions are narrow and clearly separated (except for sub-solar atmospheric metallicities). This is generally the case for low-metallicity planets, such as TOI-1259 A b. One reason for this is that the assumption $Z \geq Z_{atm}$ is enforced when the bulk metallicity is inferred, resulting in a strict lower bound and therefore a strong constraint. This separates the low $Z_{atm}$ cases, which have their peaks at low $Z$, from the higher $Z_{atm}$ ones. Notable exceptions are planets that have very low observational uncertainties; for example HATS-72 b. For most planets, there is also a clear difference between the lowest and highest atmospheric metallicities.

\begin{table*}
    \centering
    \setlength{\tabcolsep}{8pt}
    \begin{tabular}{ccccc}
        \toprule
        Planet            & Radius [R$_J$] & Mass [M$_J$] & F$_*$ [erg cm$^{-2}$ s$^{-1}$] & Age [Gyr]    \\ 
        \midrule
        HAT-P-15 b   & $1.07 \pm 0.04$ & $1.95 \pm 0.07$ & $1.5 \times 10^8$ & $6.7^{+2.7}_{-1.6}$      \\
        HAT-P-17 b   & $1.01 \pm 0.03$ & $0.53 \pm 0.02$ & $8.4 \times 10^7$ & $7.8^{+3.3}_{-3.3}$      \\
        HAT-P-18 b   & $1.00 \pm 0.05$ & $0.20 \pm 0.02$ & $1.2 \times 10^8$ & $12.4^{+2.9}_{-6.4}$     \\
        HATS-72 b    & $0.722 \pm 0.003$ & $0.125 \pm 0.004$ & $6.8 \times 10^7$ & $12.2^{+0.2}_{-0.5}$ \\
        HD 17156 b   & $1.06 \pm 0.04$ & $3.24 \pm 0.08$ & $1.4 \times 10^8$ & $3.0^{+0.8}_{-0.7}$      \\
        K2-287 b     & $0.85 \pm 0.01$ & $0.32 \pm 0.03$ & $1.0 \times 10^8$ & $4.5^{+1.0}_{-1.0}$      \\
        TOI-1259 A b & $1.02 \pm 0.03$ & $0.44 \pm 0.05$ & $1.9 \times 10^8$ & $4.8^{+0.7}_{-0.8}$      \\
        TOI-1478 b   & $1.06 \pm 0.04$ & $0.85 \pm 0.05$ & $1.6 \times 10^8$ & $9.1^{+3.1}_{-3.9}$      \\
        TOI-201 b    & $1.00 \pm 0.01$ & $0.42 \pm 0.04$ & $4.0 \times 10^7$ & $0.9^{+0.5}_{-0.5}$      \\
        WASP-69 b    & $1.06 \pm 0.05$ & $0.26 \pm 0.02$ & $1.9 \times 10^8$ & $4.5^{+7.0}_{-7.0}$      \\
        WASP-8 b     & $1.04 \pm 0.03$ & $2.13 \pm 0.08$ & $1.7 \times 10^8$ & $4.0^{+1.0}_{-1.0}$      \\
        \bottomrule
    \end{tabular}
    \caption{The 11 key targets from the Ariel mission reference sample and their radii, masses, incident stellar fluxes, and ages. For these planets, atmospheric composition measurements can lift the degeneracy in the bulk metallicity and therefore lead to a better characterisation.}
    \label{tab:key_ariel_targets}
    \setlength{\tabcolsep}{6pt}
\end{table*}

From \cref{fig:metallicity_dists_ariel} it is clear that the uncertainty over the entire range of atmospheric metallicities is much larger than for any individual case. To quantify this uncertainty, we assume that each $Z_{atm}$ is equally likely and calculate the full uncertainty as follows: Let the mean values of the bulk metallicity for $Z_{atm} = 0.5, 8 Z_\odot$ be $\mu_{Z_{0.5, 8}}$ with the corresponding standard deviations $\sigma_{Z_{0.5, 8}}$. We calculate the lower bound as $Z_l = \mu_{Z_{0.5}} - \sigma_{Z_{0.5}}$ and the upper $Z_u = \mu_{Z_{8}} + \sigma_{Z_{8}}$. The full uncertainty is then given by $\sigma_Z = Z_u - Z_l$. The corresponding mean metallicity is then simply the mean within that range.

We now compare this full uncertainty to an individual case where the atmospheric composition is known. Here, for illustration purposes and without loss of generality, we assume that each planet has its atmospheric metallicity determined as $Z_{atm} = 3 Z_\odot$. Of course in reality, the measured atmospheric metallicity will have an uncertainty that depends on the individual target and on the stellar properties. 
It should be noted that our conclusions are independent of this choice, because for each atmospheric metallicity the individual $\sigma_{Z_i}$ are nearly identical. The value was chosen to facilitate the comparison of the uncertainties since the mean bulk metallicities line up at around 3 $Z_\odot$.

\begin{figure}
    \centering
    \includegraphics[width=\columnwidth]{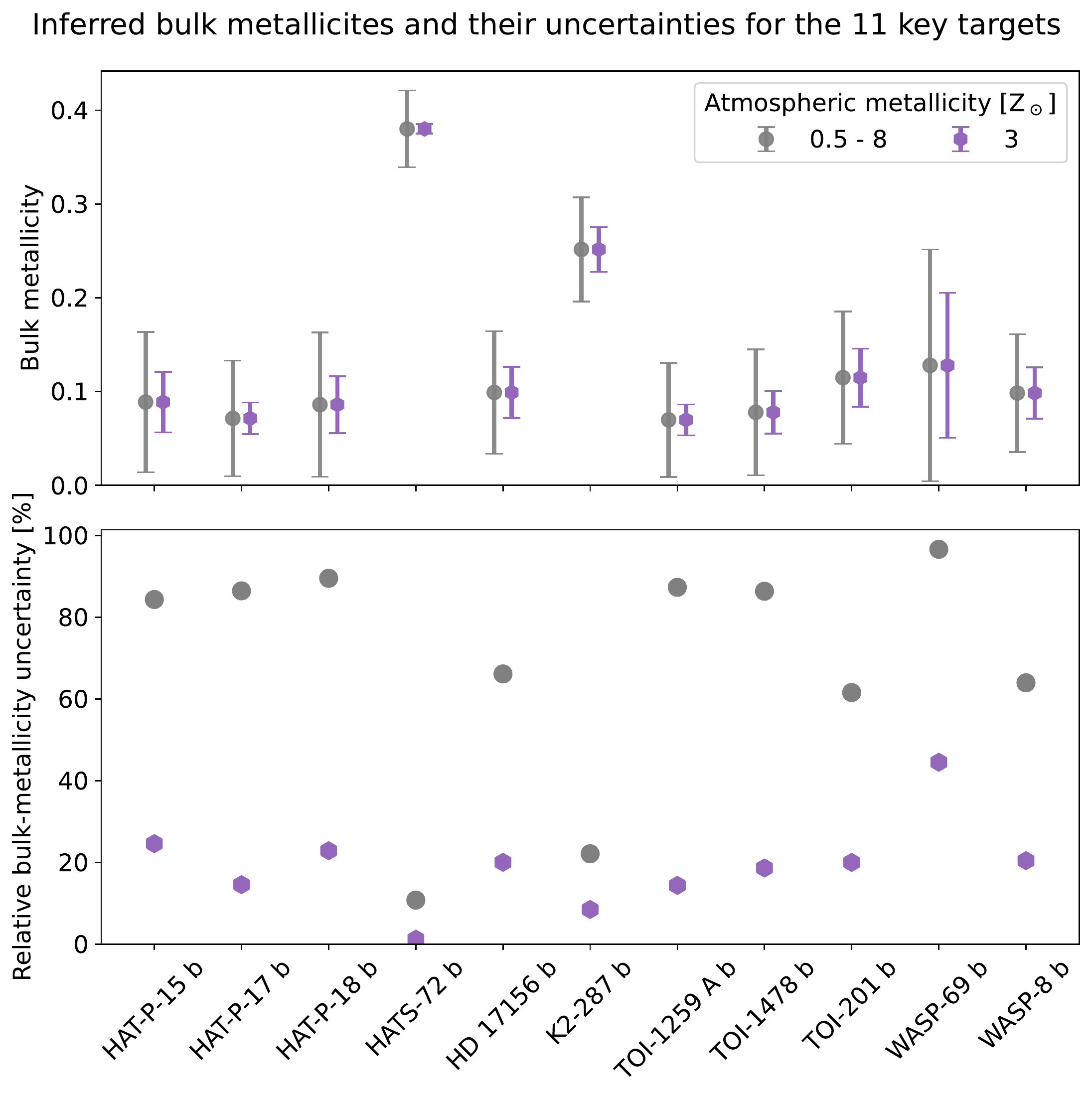}
    \caption{Top: Inferred bulk metallicities and their uncertainties for the 11 key targets from the Ariel reference mission sample. Bottom: Relative bulk-metallicity uncertainties in percent for the same planets. The grey dots and error bars represent the metallicities and uncertainties when the atmospheric metallicity is unknown between 0.5 and 8 $Z_\odot$. The purple hexagons are the results for when the atmospheric metallicity is known to be three times solar. See text for details.}
    \label{fig:uncertainty_comparison_ariel_targets}
\end{figure}

The comparison between the cases with the full uncertainty $\sigma_Z$ (grey circles) against the $\sigma_{Z_3}$ (purple hexagons) is shown in \cref{fig:uncertainty_comparison_ariel_targets}. Without information on the atmospheric composition, the relative bulk-metallicity uncertainty ($r_Z$) is mostly between 60 and 90\%. This improves drastically if $Z_{atm}$ is known, with $r_Z$ typically smaller than 30\%. In other words, the uncertainty is reduced by a factor of four to eight. In the most extreme case (HATS-72 b), the uncertainty is reduced by a factor of 16. It is therefore clear that knowledge of the atmospheric composition can significantly reduce the uncertainty in the bulk composition.

\subsection{Mass-metallicity relation}\label{sec:mass_metallicity}

We used Bayesian linear regression to infer the mass-metallicity relation for the warm gas giants from the Ariel mission reference sample (see Sect. \ref{sec:connection_to_ariel} and Appendix \ref{sec:potential_ariel_targets}). The linear regression was performed in log-space, which resulted in a power-law relation between the planetary mass $M [M_J]$ and the heavy-element enrichment of the planet $Z_{pl} / Z_{st}$, where $Z_{st} = 0.014 \times 10^{[\textrm{Fe/H}]}$ is the metallicity of the host star. The log-likelihood function for a linear model $m x + b$ is

\begin{equation}
    \begin{aligned}
        {} & \ln p(y_o, \mu_x \, | \, x, m, b, \sigma_x, \sigma_y, f) = \\
           &-\frac{1}{2} \sum_{n} = \left[\frac{\left(y_{o, n} - m x_n - b \right)^2}{\sigma_{y, n}^{2}} + \ln\left(2 \pi s_n^2\right)\right] \, ,
    \end{aligned}
\end{equation}where $x_n$ are the independent variables, $y_{o, n}$ the dependent variables, and $s_n^2 = \sigma_{y, n}^{2} + f^2 \left(m x_n + b \right)^2$ with $\sigma_{y, n}$ being the uncertainty on $y_{o, n}$. The parameter $f$ accounts for the possibility that the uncertainties $\sigma_{y, n}$ are underestimated by a (constant) fraction of $y_{o, n}$. The prior for $x$ is the (normal) distribution from observations: $p(x) \propto \prod_{n} \, \mathcal{N}(\mu_{x, n}, \sigma_{x, n})$. We used uniform non-informative priors $p(m, b, f)$ for the model parameters. The posterior probability function (up to some constant) is then

\begin{equation}
    \begin{aligned}
    {} & p(x, m, b, f \, | \, y_o, \mu_x, \sigma_x, \sigma_y) \propto \\
       & p(m, b, f) \, p(x) \, p(y_o, \mu_x \, | \, x, m, b, \sigma_x, \sigma_y, f) \, .
    \end{aligned}
\end{equation}

\begin{figure}[ht]
    \centering
    \includegraphics[width=0.75\columnwidth]{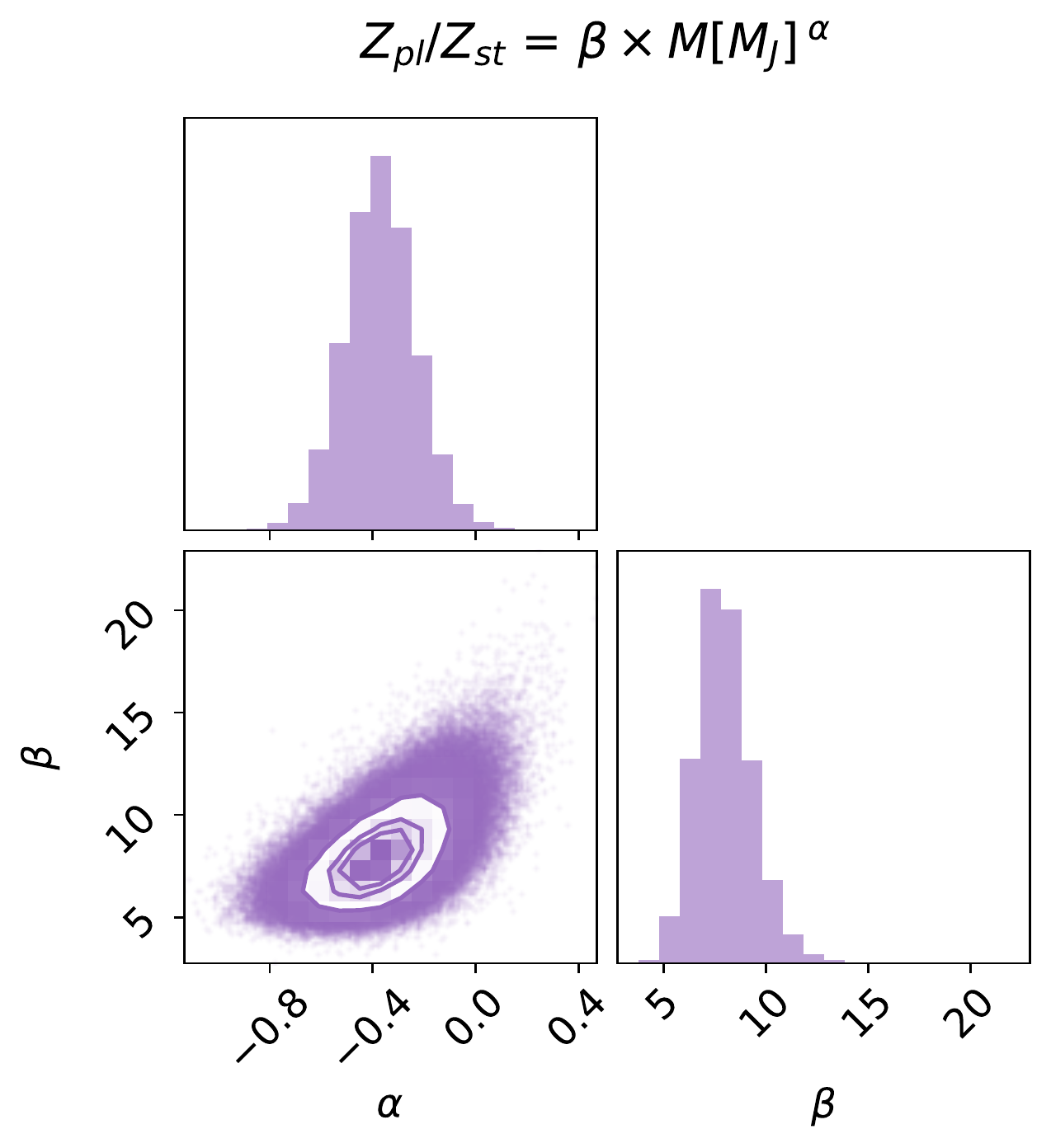}
    \caption{Corner plot showing the one- and two-dimensional projections of the posterior probability distributions of the fit parameters $\alpha$ and $\beta$.}
    \label{fig:mass_metallicity_mcmc_corner_bootstrapped}
\end{figure}

As shown in \cref{fig:metallicity_dists_ariel}, the assumed atmospheric metallicity can have a large influence on the inferred bulk composition. To account for this in our fitting procedure, we created 1,000 bootstrap samples of the planets with a random atmospheric metallicity $Z_{atm} \, [Z_\odot]$ drawn from the set $\{0.1, 0.5, 1, 2, 4, 8\}$. Since we have no information about the atmospheric composition of these planets, we assumed that they were equally likely. For each bootstrap sample, we used a Markov chain Monte Carlo (MCMC) method to estimate the posterior distribution for the fitting parameters. The combined bootstrap and MCMC samples were then used to calculate the best fit parameters. The best fit line, as well as the one-$\sigma$ error contour are shown in \cref{fig:mass_metallicity_mcmc_corner_bootstrapped}. For readability purposes, the error bars show the uncertainty of the inferred heavy-element enrichment for a solar metallicity atmosphere. \cref{fig:mass_metallicity_mcmc_corner_bootstrapped} shows the two-dimensional projection of the posterior probability distributions of the power-law $Z_{pl} / Z_{st} = \beta \times M [M_J] \, ^{\alpha}$ fit parameters $\alpha \equiv m$ and $\beta \equiv 10^b$.

\begin{figure}[ht]
    \centering
    \includegraphics[width=\columnwidth]{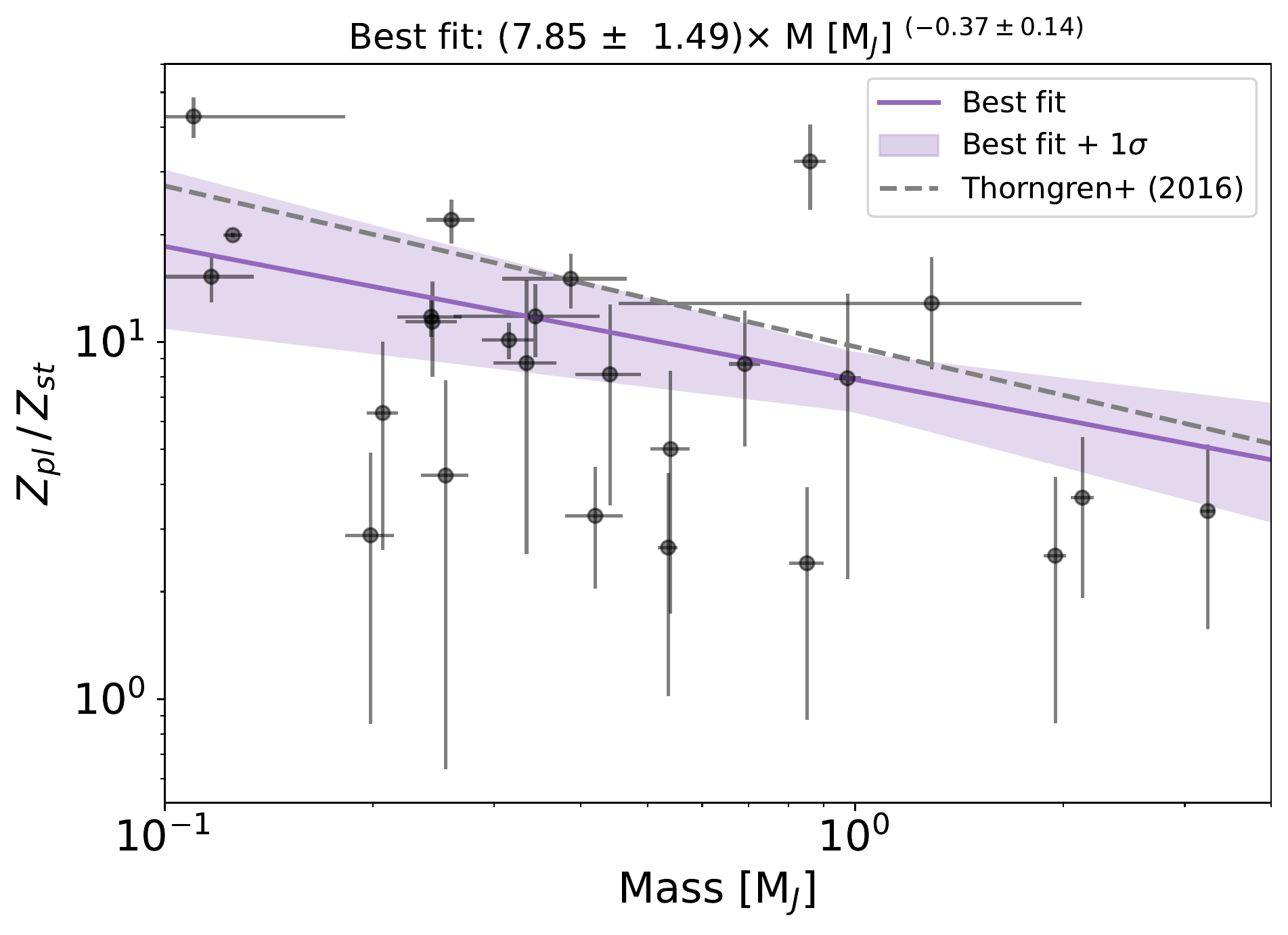}
    \caption{Heavy-element enrichment as a function of the planetary mass. The black dots with error bars show the inferred metallicity of the potential Ariel targets. The solid purple line shows the best-fit line, the purple-shaded region shows the 1$\sigma$ error contour around the best fit, and the dashed grey line shows the best fit from \citet{Thorngren2016}.}
    \label{fig:mass_metallicity_mcmc_bootstrapped}
\end{figure}

The best fit parameters are $\alpha = -0.37 \pm 0.14$ and $\beta = 7.85 \pm 1.50$. We calculated Kendall's tau correlation and the corresponding p-value from the bootstrap sample, which yielded a correlation of -0.32 (p-value: 0.035), where the p-values of the bootstrap samples were calculated as described in \citet{efron1994introduction}. Since we were using bootstrap samples, the correlation and the p-value accounted for the uncertainties in the independent (planetary mass) as well as the dependent (inferred metallicities) variables. 

For comparison, \citet{Thorngren2016} inferred $\alpha_{T16} = -0.45 \pm 0.09$, $\beta_{T16} = 9.70 \pm 1.28$ and a correlation of $-0.44$ (p-value: $1.3 \times 10^{-5}$) for the planets in their sample. They also calculated Kendall's tau; however they did not account for the uncertainties and used the mean values instead. While similar, our best fit for the Ariel sample yields a lower metal-enrichment at a given planetary mass, drops off more slowly (flatter slope) and is more uncertain. However, the \citet{Thorngren2016} best fit is mostly consistent within one $\sigma$ of our results. 

The lower metal-enrichment trend can be explained by the different hydrogen-helium equations of state that were used. Here, the evolution models used the \citet[][hereafter CMS]{Chabrier2019} hydrogen-helium equation of state, while \citet{Thorngren2016} used the \citet[][hereafter SCvH]{Saumon1995} one. It has been shown previously that CMS generally leads to smaller planetary radii, and therefore a lower heavy-element content will be inferred for a given observed planet \citep{2020ApJ...903..147M}. On the other hand, the different slope and the higher uncertainties of the fit parameters are likely the results of our models considering non-solar atmospheres, and having fewer (and a few different) planets for the fit. These are also the likely causes for the lower and less statistically significant correlation.

As discussed in \citet{Thorngren2016}, these results are inconsistent with planetary population synthesis models \citep[e.g.][]{Mordasini2014}. \citet{Hasegawa2018} suggested that a power law with roughly $Z_{pl} / Z_{st} \propto M [M_J] \, ^{-0.4}$ emerges if the metal enrichment is a tracer for the final stages of giant planet formation, and the heavy elements are mostly accreted from gapped planetesimal disks. This interesting topic should be investigated further by developing more comprehensive formation models and their comparison with future accurate data.

\section{Radius inflation due to collisions}\label{sec:collisions}

\addone{Our models assumed that the planet is isolated (except for the irradiation). Therefore, we did not account for events such as giant impacts, which could affect the subsequent evolution and therefore the observed radius. While no direct observations exist, theoretical studies suggest that giant impacts are common during the early evolution of planetary systems \citep[e.g.][]{2015MNRAS.446.1685L}. Collisions have also been used to explain, for example, Jupiter's apparently dilute core \citep{Liu2019}, the high metallicity of some exoplanets \citep[e.g.][]{2021A&A...648L...1O} or their eccentricities \citep{2019ApJ...884L..47F}. Typically, evolution models do not include the effects such impacts could have on the predicted sizes of exoplanets.}

\addone{Here, we investigated how the deposited energy from collisions inflates the radius of a giant planet, and the timescale associated with that inflation. Our approach was as follows: We started with a somewhat typical warm giant ($M = 1 M_J$, $Z = 0.1$, $F_* = 10^7$ erg cm$^{-2}$ s$^{-1}$), and let it evolve for 2 Gyr (since most observed warm giants are commonly at least this old). To simulate an impact, we injected a certain amount of energy into the giant planet on a very short timescale, and ignored the mass deposition for simplicity. The specific energy of the impact (erg/g) was deposited uniformly between a pressure of $P_{top} = 10^6$ Ba and $P_{bot}$, which we varied from $10^9 - 10^{13}$ Ba or used the pressure at the centre. This accounts for the fact that the impactor would likely be ablated and ripped apart by the dense envelope of the target. Afterwards, the planet evolved and cooled usual.}

\addone{Guided by the simulations of the giant impacts for Jupiter of \citet{Liu2019}, we considered impact energies of $E_i = 10^{38}, 10^{39}, 10^{40}, 10^{41}$, and $10^{42}$ erg. The upper limit roughly corresponds to the difference in energy of Jupiter before and after the collision with a $10 M_\oplus$ planetary embryo in \citet{Liu2019}, and is similar to the impactors kinetic energy. If the impact velocity were roughly at the escape speed of Jupiter today, this would correspond to the kinetic energy of the impact with approximate masses of $M_i \simeq 10^{-3}, 10^{-2}, 10^{-1}, 1,$ and $10$ $M_\oplus$. Clearly such impacts would also alter the planetary interior structure, deposit mass and could also cause envelope mass-loss. We also did not consider that a very energetic impact could create composition gradients and inhibit convection, leading to a non-adiabatic evolution. Instead, we intentionally kept our model simple in order to focus on studying the effect of the post-impact radius inflation that is solely associated with the energy of the impact.} 

\begin{figure}[ht]
    \centering
    \includegraphics[width=0.7\columnwidth]{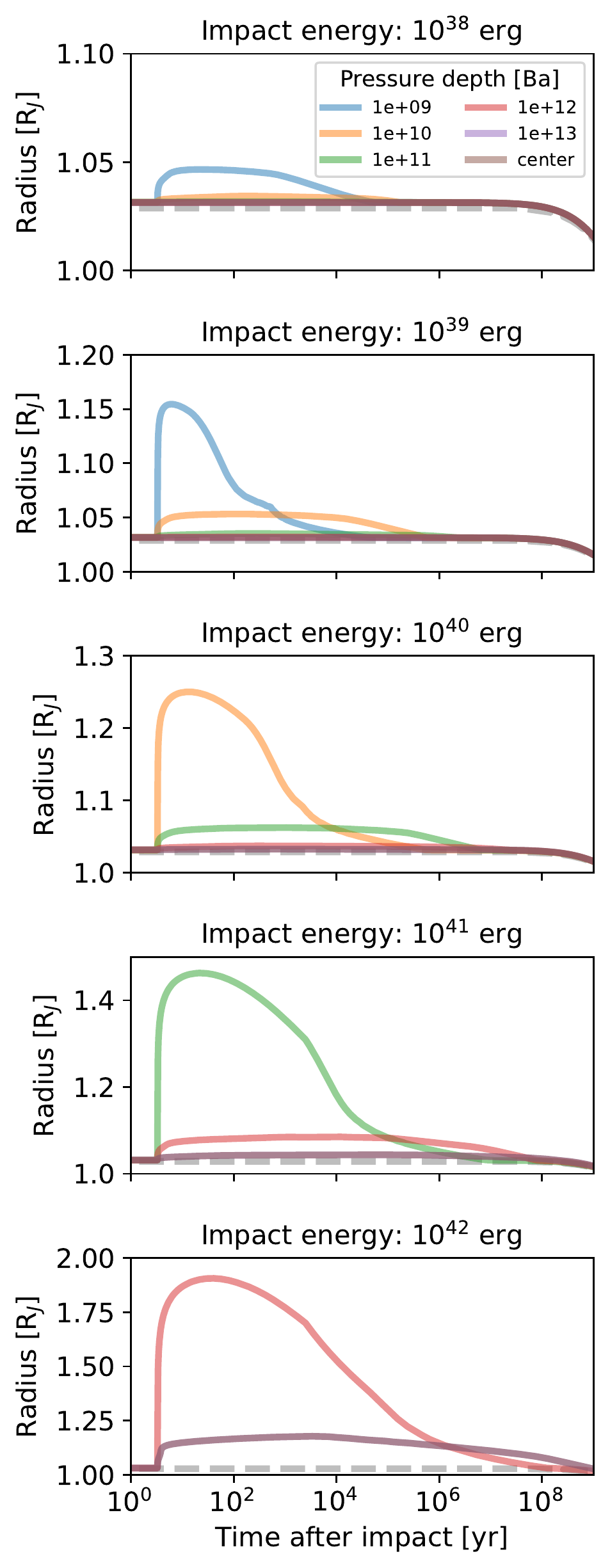}
    \caption{Radius evolution after the impact. The rows show the evolution for different impact energies. The colours correspond to the upper pressure boundary (lower boundary in terms of radial coordinates) of where the energy is deposited (see text for details). The dashed grey line shows the radius evolution without an impact. Some coloured lines for specific combinations of the impact energy and pressure depth are missing because they resulted in non-hydrostatic configurations.}
    \label{fig:radius_after_impact}
\end{figure}

\addone{\cref{fig:radius_after_impact} shows the radius evolution after the impact for all the cases we considered. Each row shows a different impact energy, and the colours correspond to the various $P_{bot}$ values. For impact energies beyond $10^{39}$ erg, using small $P_{bot}$ values resulted in a non-hydrostatic configuration, and therefore the evolution could not be simulated further. It is unlikely that energetic impacts are associated with impactors that only penetrate to such low pressures (see e.g. the $1 M_\oplus$ impact of \citet{Liu2019}). Depending on the impact energy and $P_{bot}$, shortly  after the impact the giant planet can be significantly inflated, up to almost twice of its original size (for the highest energy we considered). However, the post-impact cooling is relatively rapid in most cases. We find that for the radius to remain inflated beyond a few  times$10^3 - 10^4$ years, either $P_{bot}$ has to be small or the impact must be highly energetic, requiring very massive impactors that are unlikely to be present at ages of Gyrs. We do, however, find that impacts could affect the early evolution of young giant planets. This could affect the interpretation of directly imaged planets and we hope to address this in future research.}

\section{Discussion}\label{sec:discussion}

The characterisation of giant exoplanets relies on evolution simulations that require assumptions and simplifications. The limitations of the evolution models used in this work are discussed in detail in \citet{2021MNRAS.507.2094M}. Here, we briefly summarise some of the most influential assumptions and parameters. Important factors are the equations of state \citep{Baraffe2008,Vazan2013}, atmospheric model \citep{Burrows2007,Poser2019,2020ApJ...903..147M}, the distribution of heavy elements in the planetary interior \citep{Baraffe2008,Thorngren2016}, and whether the interiors are fully convective \citep{Leconte2012,Kurokawa2015}. These parameters are not fully constrained and are therefore limitations of evolution models in general. In most cases, they can influence the predicted radius on the percent level, which is larger than some observational uncertainties today.

\addone{An important uncertainty is whether grains exist in the atmospheres of exoplanets, which could slow the planetary cooling due to an enhancement of the opacity \citep{Movshovitz2010,Vazan2013,2014A&A...566A.141M}. While our evolution models do not include the grain opacity, we briefly demonstrate how the cooling could be affected in Appendix \ref{sec:appendix_grain_opacity}. If grains cause the atmosphere to be significantly more opaque, the inferred metallicities would be shifted towards higher values.}

\addone{An additional possible mechanism for changing the radius of a giant exoplanet is a giant impact, which we have investigated in Sect. \ref{sec:collisions}. It should be noted that collisions may be unique to any given planet, and that the possible characteristics of potential collisions include a very large parameter space (mass of the impactor, impact timing, impact velocity, etc.), which is difficult to generalise. While the aim of this work is to investigate exoplanets on a population level, it is nonetheless valuable to estimate the degree to which collisions leave a planet inflated. Our results suggest that only when collisions are extremely energetic ($> 10^{41}$ erg) a significant inflation lasting longer than a few thousand years is expected. Unless giant impacts are very frequent or extremely energetic, which is unlikely for older planetary systems \citep{2015MNRAS.446.1685L,2019ApJ...884L..47F}, collisions are unlikely to significantly inflate giant exoplanets over long timescales.}

For planets younger than a few tens of megayears, additional uncertainties are the unknown initial thermal state of the planets \citep[e.g.][]{2003A&A...402..701B,Marley2007,Berardo2017a} and the fact that the evolution of the host star is not considered. 
Since in this work we focus on older planets, these uncertainties are relatively unimportant. 

Another important source of uncertainty is linked to the commonly used hydrogen-helium equations of state \citep[e.g.][]{Saumon1995,Chabrier2019} that do not include the non-ideal interactions of the two elements. Recently, \citet{2021ApJ...917....4C} showed the importance of this effect in the low-temperature and high-density conditions expected in giant planets (and brown dwarfs). They provided a new effective hydrogen table, which can be used to calculate hydrogen-helium mixtures including these interactions. It is, however, unclear to what extent this would influence the mass-radius relation of giant planets. While the implementation of this new equation of state into the evolution models is beyond the scope of this work, we calculated hydrostatic models of pure hydrogen-helium planets in order to quantify this effect. We find that for planets with masses between 0.1 - 10 $M_J$, the inclusion of the non-ideal effects from \citet{2021ApJ...917....4C} yielded a 1-2\% larger radius compared to the \citet{Chabrier2019} equation of state. Therefore, this uncertainty is comparable in magnitude to the other model assumptions.

In this work we did not focus on the characterisation of giant exoplanets but rather on the conditions for which more precise or additional measurements can lead to lower uncertainties in the bulk composition. Our calculations depend on the above assumptions: Changing, for example, the equation of state would lead to a systematic shift in the bulk metallicities. It would, however, not change the uncertainty of the result for a given observation and theoretical model. Therefore, this does not affect our main conclusions.

In Sect. \ref{sec:connection_to_plato} we investigated how the uncertainty in the bulk composition depends on the uncertainty of the stellar age measurement. In order to isolate this variable, uncertainties in the mass and radius had to be a fixed value. We set this value to 1\%. The reason behind this choice was that larger uncertainties, in particular for the radius, would quickly be the main contributors to the bulk-metallicity uncertainty. Currently, there is not a large number of warm giant exoplanets with similar measurement errors (see \cref{fig:error_dists_full} and \cref{fig:error_dists_sample}). However, this is expected to change in the near future, with more precise mass-radius measurements from ground missions as well as the Plato mission. We also set the mass and radius of the synthetic planets that we have investigated to $R = 1 R_J$ and $M = 1 M_J$. While this choice is biased towards a Jupiter analogue, it is a reasonable representation of a somewhat typical warm gas giant (see \cref{fig:planets_sample}). We performed additional calculations with different masses and radii, and confirmed that they lead to the same conclusions.

In Sect. \ref{sec:connection_to_ariel} we found that there are currently only 27 warm giants with measured masses and radii that are part of the Ariel mission reference sample. Detections of hot Jupiters are much more numerous, and the number of observed warm giants is comparably small. This is clearly not ideal for statistical inferences, such as constructing a mass-metallicity relation. Hopefully, in the near future new observations will increase the number of observed cool and warm giants substantially, which will make statistical inferences more reliable.

\addone{We note that in our models the atmospheric metallicity is used as a single free parameter, which is clearly a simplification. In reality, the metallicity is made up of different chemical elements present in various ratios, influenced by the composition of the host star and the formation history of the planet \citep[e.g.][]{2018ExA....46...45T}. In addition, exo-atmospheric observations determine the abundance of specific compounds at relatively high regions in the atmosphere and not the overall atmospheric and planetary bulk metallicity. It takes an additional, model-dependent step to infer the overall atmospheric metallicity \citep[e.g.][]{2018RNAAS...2..128H}. Many of the current atmospheric composition measurements are limited to a narrow wavelength, and cannot measure the abundance of all the chemical species that carry the most common heavy elements \citep{2019PASP..131i4401Z}. The Ariel spacecraft is expected to have a broad spectral and molecular coverage, which should provide much better estimates of the overall atmospheric metallicity \citep{Tinetti2018}. We suggest that detailed investigations of the connection between the compositions of the outer atmosphere, the deeper atmosphere and the bulk interior should be investigated in future work.}

The results from Sects. \ref{sec:connection_to_plato} and \ref{sec:connection_to_ariel} suggest that the bulk-metallicity uncertainty due to not knowing the atmospheric composition is larger than the one related to the stellar age uncertainty. This highlights the great power of atmospheric characterisation in determining the composition of giant exoplanets, and also shows that there is a synergy between more accurate age and atmospheric measurements. While our focus in this work was on the planets from the Ariel mission reference sample, our findings are also valid for observations with JWST.

\section{Summary and conclusions}\label{sec:conclusions}

In this paper we have studied in detail how the accurate determination of stellar ages from Plato and atmospheric measurements from Ariel will improve the characterisation of giant exoplanets. Our main results can be summarised as follows:

\begin{enumerate}
    \item The accurate determination of stellar ages from the Plato mission will significantly constrain the bulk composition of young to intermediate-aged planets (below a few gigayears) and reduce the uncertainty by up to a factor of two. This is true for both weakly and highly irradiated warm giants. 
    \item The Ariel mission reference sample includes 27 warm gas giants with parameters that would allow an accurate determination of their bulk compositions. 
    \item Our Bayesian linear regression resulted in the following power-law relation between the planetary mass and metal enrichment for these 27 warm giants: $Z_{pl} / Z_{st} = (7.85 \pm 1.50) \times M [M_J] \, ^{-0.37 \pm 0.14}$. This suggests a lower metal enrichment at a given planetary mass compared to previous studies.
    \item We find that the observed radius of TOI-1899 b is  $\sim$15\%  larger than values from evolution models. We therefore suggest that it must be inflated by a mechanism that is not considered in standard evolution models.
    \item Measuring the atmospheric metallicity of giant planets can reduce the uncertainty in the bulk composition by a factor of about four to eight. The largest improvements are expected for planets that have low radius, mass, and age measurement uncertainties  (in this order).
    \item We identify 11 high-priority warm giants from the Ariel mission reference sample for atmospheric characterisation (see \cref{fig:metallicity_dists_ariel} and \cref{tab:key_ariel_targets}). For these planets, measuring the atmospheric metallicity will break the degeneracy in their inferred bulk compositions.
    \item \addone{We find that impacts are unlikely to cause old giant planets to be significantly inflated. However, they could influence the early evolution of young giant planets, which could affect the interpretation of directly imaged planets.}
\end{enumerate}

Our study shows the important role of evolution models in interpreting the upcoming accurate measurements from space missions dedicated to giant exoplanet detection and characterisation. It is clear that there is a great synergy between atmospheric measurements and a more accurate determination of planetary masses, radii, and stellar ages, and that their fusion will considerably improve our understanding of  giant exoplanets.

\clearpage
\begin{acknowledgements}
    \addone{We thank the anonymous referee for thoroughly reading our paper and providing useful comments and suggestions.}
    We gratefully acknowledge support from the UZH Forschungskredit grant \texttt{FK-21-132}, the SNSF grant \texttt{\detokenize{200020_188460}} and the National Centre for Competence in Research ‘PlanetS’ supported by SNSF. This research used data from the NASA Exoplanet Archive, which is operated by the California Institute of Technology, under contract with the National Aeronautics and Space Administration under the Exoplanet Exploration Program. 
    \addone{Extensive use was also made of the MESA code \citep{Paxton2011,Paxton2013,Paxton2015,Paxton2018,Paxton2019,2022arXiv220803651J} and the python packages NumPy \citep{harris2020array}, SciPy \citep{2020SciPy-NMeth}, Matplotlib \citep{Hunter2007}, Jupyter \citep{jupyter}, emcee \citep{2013PASP..125..306F} and planetsynth \citep{2021MNRAS.507.2094M}.}
\end{acknowledgements}

\bibliographystyle{aa}
\bibliography{library}

\appendix

\section{Potential Ariel targets}\label{sec:potential_ariel_targets}
\begin{figure}[ht]
    \centering
    \includegraphics[width=\columnwidth]{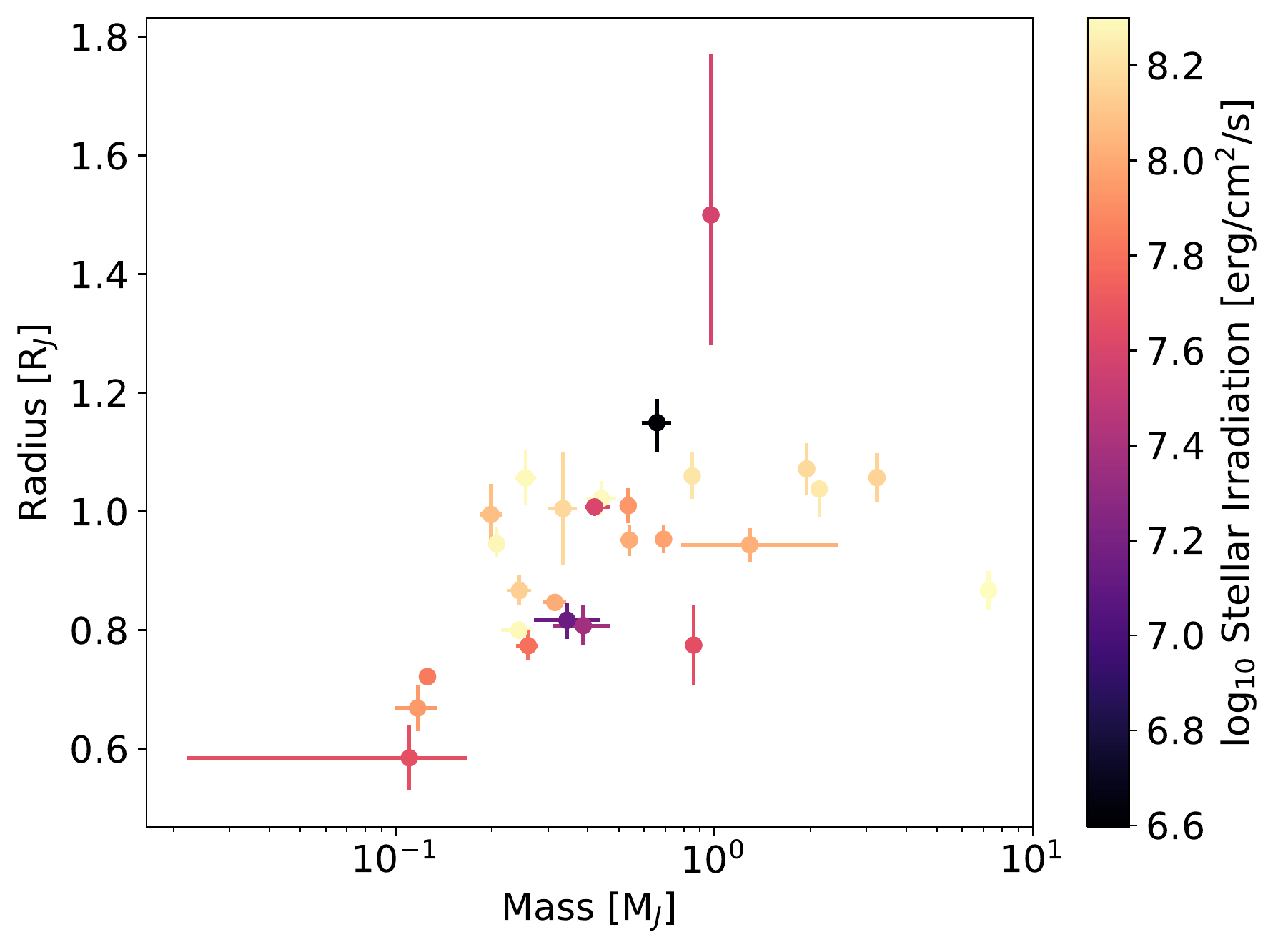}
    \caption{Mass-radius relation of the planets in the sample. The colours indicate the stellar irradiation received by the planets.}
    \label{fig:planets_sample}
\end{figure}

As described in Sect. \ref{sec:connection_to_ariel}, we selected the warm giants ($F_* \leq 2 \times 10^8$ erg cm$^{-2}$ s$^{-1}$) with measured masses, radii and stellar ages from the Ariel mission reference sample. We also limited their masses in the range of $0.1 - 10 M_J$. These planets are shown in the mass-radius plane in \cref{fig:planets_sample} coloured according to their incident stellar irradiation. It can be seen that the planets occupy the full mass range, and for any given mass there are planets with different radii.

\begin{figure}[ht]
    \centering
    \includegraphics[width=\columnwidth]{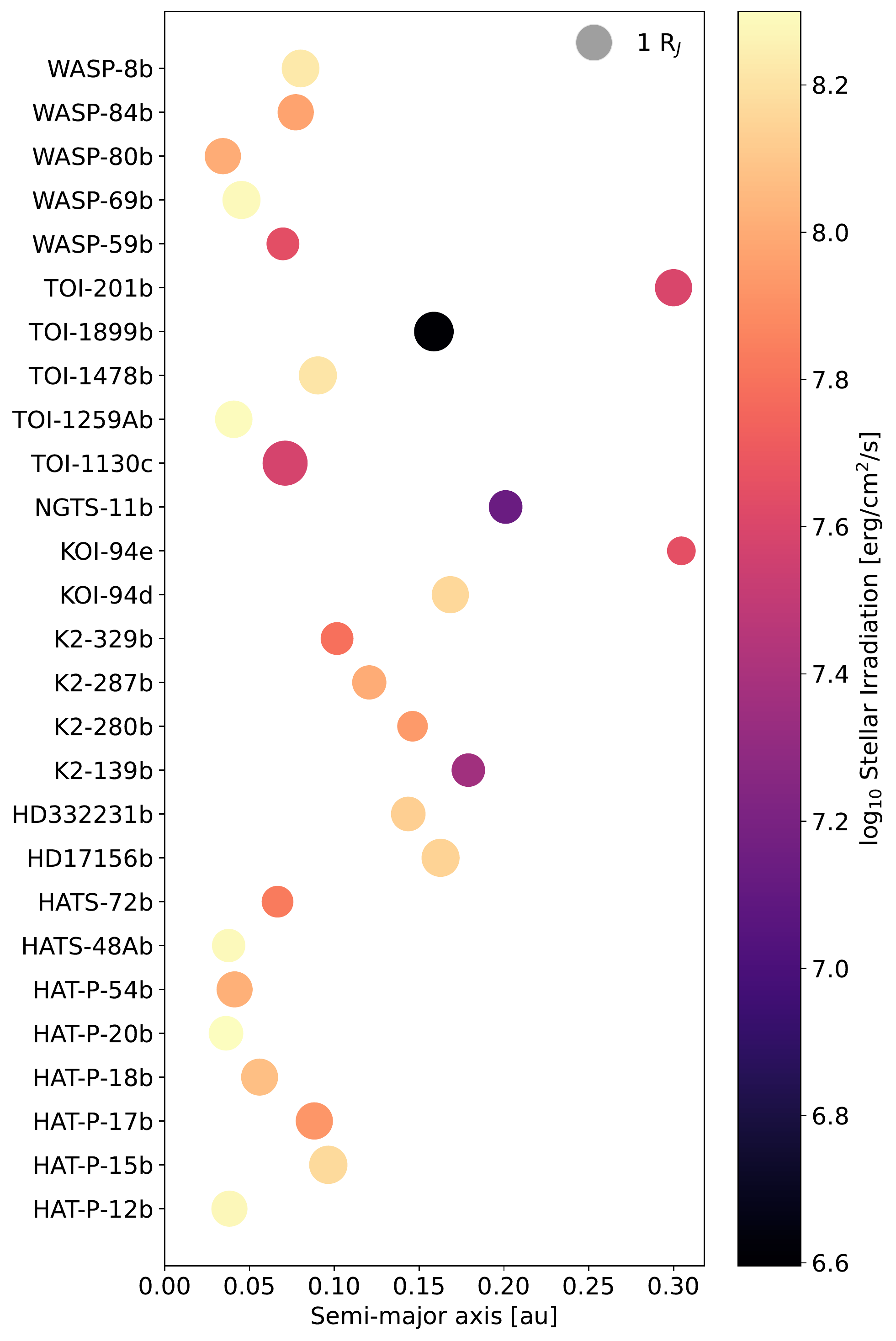}
    \caption{Planets in the sample as a function of their semi-major axis. The size of a dot corresponds to the planet's radius, and the colour shows its incident stellar irradiation.}
    \label{fig:planets_sample_listed}
\end{figure}

We also list the warm giants in \cref{fig:planets_sample_listed} together with their semi-major axis. It is clear that all the warm giants in the Ariel mission reference sample have tight orbits around their host star. None of them are cool giants.

\section{Observational uncertainties of giant planets}\label{sec:observational_uncertainties}

\begin{figure}[ht]
    \centering
    \includegraphics[width=\columnwidth]{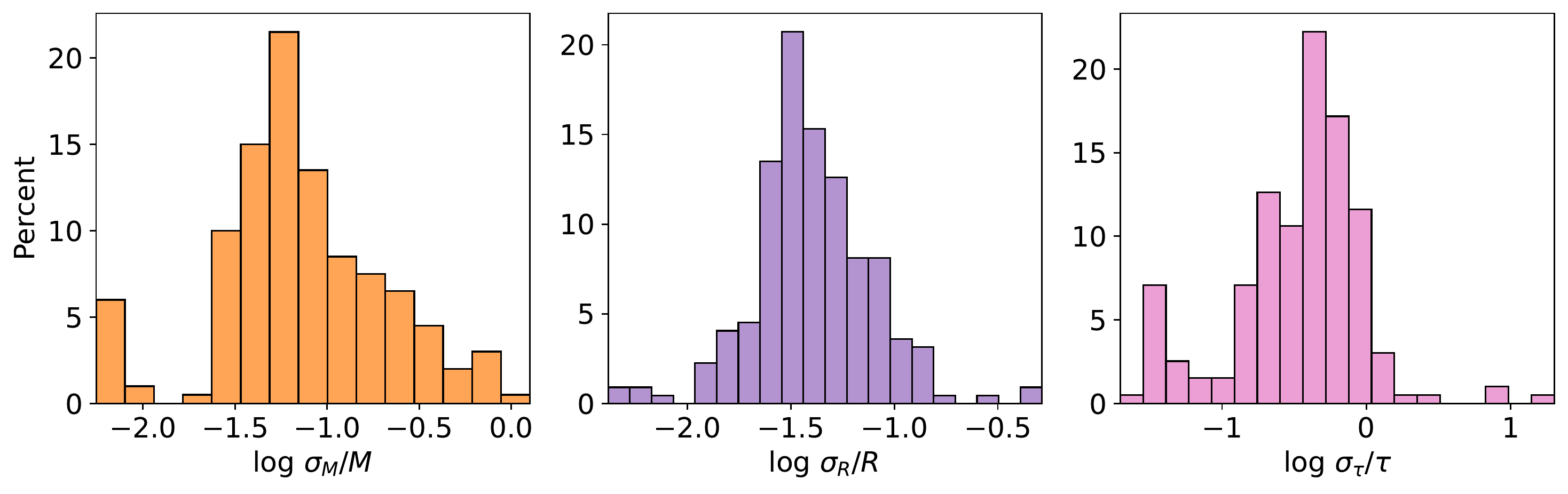}
    \caption{Histograms of the relative measurement uncertainties in mass (left), radius (centre), and age (right) of observed giant exoplanets with $0.1 < M < 10 M_J$. The data are taken from the NASA Exoplanet Archive.}
    \label{fig:error_dists_full}
\end{figure}

As discussed throughout this work, the measurement uncertainties of the planetary and stellar parameters have a large influence on the inferred bulk-metallicity uncertainty. In \cref{fig:error_dists_full}, we show the distributions of the relative measurements uncertainties ($\sigma_x / x$) for the planetary masses and radii, as well as the stellar ages for all giant exoplanets with masses of $0.1 \leq M [M_J] \leq 10$. This population includes hot Jupiters, which are currently unavailable for characterisation due to their inflated radii. The relative uncertainties peak at around 5\% for the mass, 3\% for the radius and 50\% for the stellar age.

We show the same distributions for the potential Ariel targets (see \cref{fig:planets_sample_listed}), which does not include hot Jupiters. Compared to \cref{fig:error_dists_full}, these planets have higher observational uncertainties for the mass, similar for the radius and lower for the stellar age.

\begin{figure}[ht]
    \centering
    \includegraphics[width=\columnwidth]{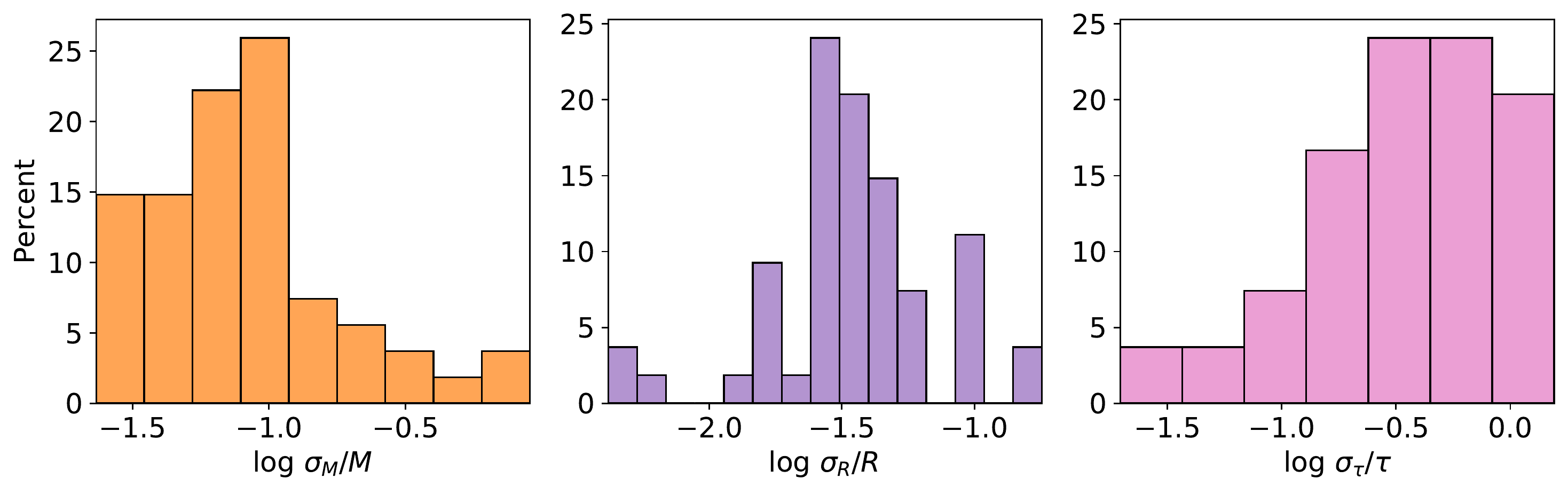}
    \caption{Histograms for the relative uncertainties in mass (left), radius (centre), and age (right) of the potential Ariel targets (see Fig. \ref{fig:planets_sample_listed}).}
    \label{fig:error_dists_sample}
\end{figure}

\section{The influence of grain-enriched atmospheres}\label{sec:appendix_grain_opacity}

\addone{While the opacity used in this work accounts for the influence of metals, we did not consider an opacity enhancement due to grains. Grain-rich atmospheres are more opaque, effectively trapping heat from the interior for longer, particularly at young ages \citep{Vazan2013}. Currently it is still unclear whether grains are present in exoplanet atmospheres over large timescales, or settle quickly to deeper regions \citep{Movshovitz2010,Mordasini2014}.}

\addone{Here, we investigated the effect of grains on the cooling of giant exoplanets. We calculated the evolution of a $M = 1 M_J$, $Z = 0.1$ exoplanet with a modified version of MESA that includes a grain opacity \citep{2020ApJ...903..147M,2021MNRAS.507.2094M}. We used the analytical fit from \citet{Valencia2013} to the interstellar medium (ISM) grain opacity from \citet{Ferguson2005}. ISM grains are very small, and therefore strongly enhance the opacity. However, the full ISM grain opacity is likely about two orders of magnitude too large for exoplanets \citep{Movshovitz2010,Mordasini2014}. Therefore, we calculated the evolution several times using the full, 10\%, 1\% and no grain opacities. Since the grain opacity is temperature dependent, we also used two different stellar irradiation fluxes of $10^4$ and $10^7$ erg cm$^{-2}$ s$^{-1}$, which are typical fluxes for cold and warm Jupiters.}

\begin{figure}[ht]
    \centering
    \includegraphics[width=\columnwidth]{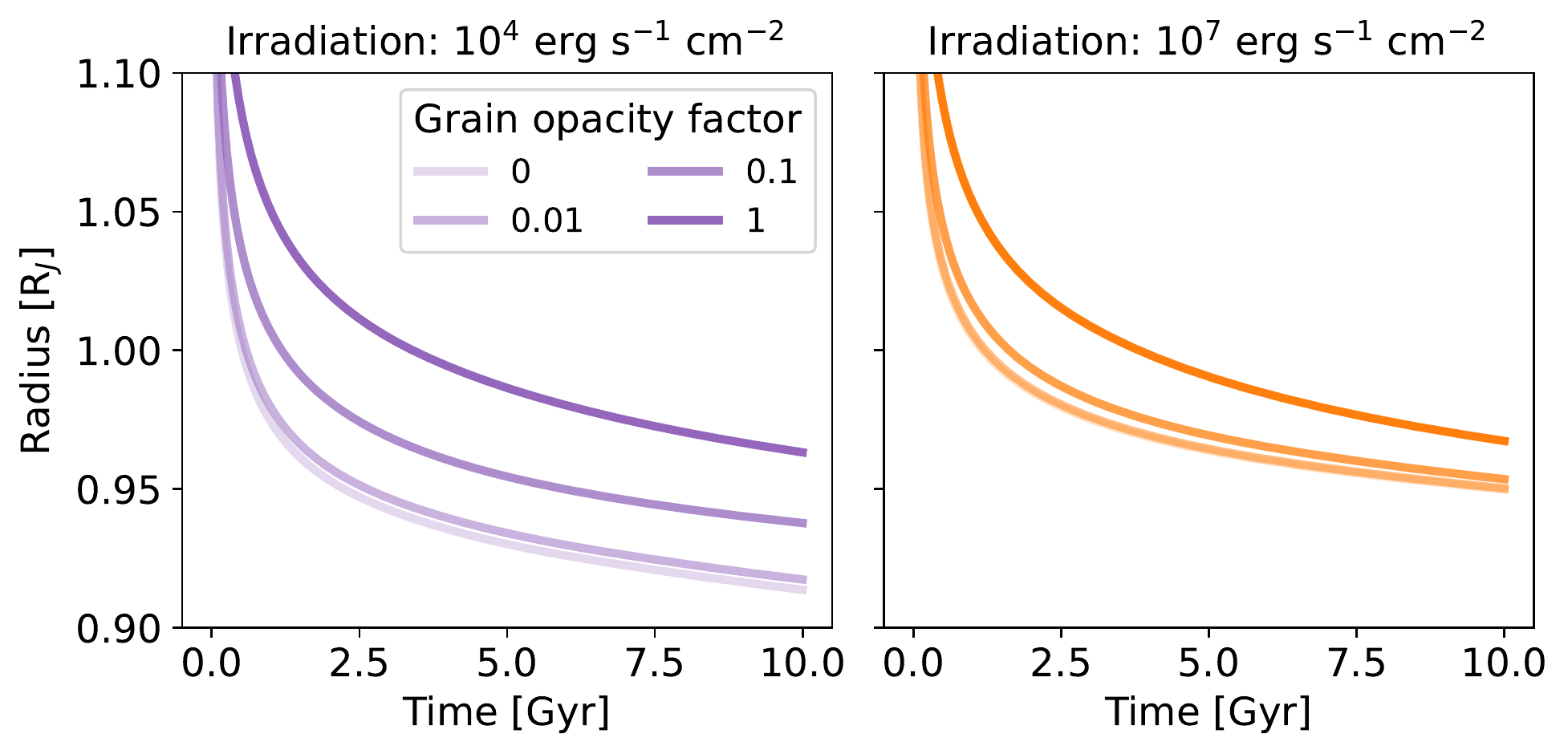}
    \caption{Radius evolution for a $M = 1$ M$_J$, $Z = 0.1$ planet with different grain opacities (see the legend) and stellar irradiation fluxes of $10^4$ (left) and $10^7$ erg cm$^{-2}$ s$^{-1}$ (right).}
    \label{fig:radius_grain_opacity}
\end{figure}

\addone{In \cref{fig:radius_grain_opacity} we show the radius evolution for all cases. With both stellar fluxes, the full ISM grain opacity yields significantly larger radii, even at an age of a few gigayears. When the grain opacity is reduced by a factor of ten, the inferred radii are still significantly larger for the low-flux case, while for the high-flux case only a small inflation remains. Once the grain opacity is reduced to 1\% of its original value, the radius of the low-flux planet is very similar to the case without grain opacity.}

\addone{A strongly enhanced opacity would lead to larger radii at a few gigayears, and therefore change the inferred bulk metallicity. As noted above, in exoplanets the grain opacity is likely to be $\lessapprox 1\%$ of the ISM grain opacity, and therefore the effect is expected to be small. There are other possible way to enhance the opacity, such as thick cloud layers \citep{Poser2019}. However, these cloud models have many free parameters that are not well constrained. It is therefore important to develop realistic atmospheric models that include grains and clouds in order to take full advantage of future observations.}

\end{document}